# On Features of Scaling in Duffing Oscillator Under Action of Impulses with Random Modulation of Parameters

Alexander P. Kuznetsov [1, 2], Julia V. Sedova [1, 2]

[1] *Institute of Radio-Engineering and Electronics of RAS, Saratov Division, Saratov, Russian Federation*

[2] *Saratov State University, Saratov, Russian Federation*

**Abstract**

In the work a nonlinear Duffing oscillator is considered under impulse excitation with two ways of introduction of the random additive term simulating noise, - with help of amplitude modulation and modulation of period of impulses sequence. The scaling properties both in the Feigenbaum scenario and in the tricritical case are shown.

## 1. Introduction

Some basic scenarios of transition from regular behavior to chaos for nonlinear systems are known. All of them allow description with the help of a renormalization group method and therefore have property of universality concerning a concrete kind of system [1]. One of consequences of an opportunity of a renormalization approach is the property of scaling - reproducibility in more fine scales of the dynamics characteristics in the vicinity of a critical point of transition to chaos. The scaling property can be observed in the structure of bifurcation tree, Lyapunov exponent diagram, phase portrait of attractor, etc. In real systems, however, always noise is presented, which can greatly influence at the picture of transition. For example in a case of period-doublings the noise "washes away" thin structure of bifurcation tree, as a result only the limited number of period-doublings can be observable (so much the greater the less a noise intensity). Thus the known illustrations of scaling, generally speaking, become impossible, because they assume as much as close approach to a critical point. Uncommonly, that the renormalization group method can be generalized on a case of systems with noise (see for example, [2-7]). Due to this it is possible to distribute on stochastic systems the scaling property. For that at transition from one level of hierarchy to another we should to renormalize noise amplitude on new universal constants. For a number of universality classes such constants are determined [2, 5-8]. At the same time it is important to give scaling illustrations not only for formal "canonical" models, but also for maps obtained from "the first principles" for physical systems, when the connection of parameters with parameters of initial physical system is determined. Besides it would be interesting to present illustrations of self-similarity property in systems with period-doublings under action of noise on a plane of essential parameters of system. In work in this context nonlinear Duffing oscillator with periodic impulse influence is studied. The noise is brought into system as random modulation either impulses amplitudes, or intervals of time between the neighboring impulses. It allows to obtain one-dimensional maps analytically.

We shall consider two classes of universality connected with period-doublings. First one is a classical Feigenbaum behavior type and is characterized, together with Feigenbaum constants $\delta_F$ = 4.669201609… and $\alpha_F$ = -2.502907875.., by noise constant $\mu_F$ = 6.61903.., found out in works of Crutchfield et al. and Shraiman et al. [2, 8]. The second type of behavior assumes generalization of the Feigenbaum scenario at a case of two-parametrical maps. It is characterized by the universal self-similar structure of a system's parameters plane in a vicinity of terminal points for Feigenbaum lines - tricritical points. This organization for systems without noise was discussed in [9] with the help of charts of dynamical regimes. We shall demonstrate the scaling regularities with the help of investigating of two-dimensional picture of Lyapunov exponent distribution both in systems without noise, and with noise. We shall specify the appropriate universal constants: $\delta_T$ = 7.284686217.., $\alpha_T$ = -1.69030297… and noise constant $\mu_T$ = 8.2439...

**2. Investigated system**

So, we shall consider nonlinear Duffing oscillator with periodic impulse influence. Let us assume, that the action of impulses takes very short time, so for this time coordinate of oscillator practically has no time to change, and the speed obtains addition determined by impulse amplitude. The behavior of such oscillator is described by the following differential equation:

$$\ddot{X} + \gamma\dot{X} + \omega_0^2 X + \beta X^3 = \sum C\delta(t - nT), \qquad (1)$$

where $X$ - dynamical variable, $\gamma$ - damping factor, $\omega_0$ - eigen frequency, $\beta$- parameter of nonlinearity, $T$ - period of external impulses sequence, $C$ – impulse amplitude.

Because in a right part of the equation (1) there is a $\delta$-function, then in interval between impulses the right part of the equation makes vanish:

$$\ddot{X} + \gamma\dot{X} + \omega_0^2 X + \beta X^3 = 0. \qquad (2)$$

In this case it is possible to obtain an approached analytical solution, using a method of slowly varying amplitudes. Within this method we shall present coordinate $X$ as:

$$X = \frac{a}{2}e^{i\omega_0 t} + \frac{a^*}{2}e^{-i\omega_0 t} \qquad (3)$$

where $a(t)$ and $a^*(t)$ – complex and complex conjugate slowly varying amplitudes correspondingly. We shall take into account a traditional additional condition

$$\dot{a}e^{i\omega_0 t} + \dot{a}^*e^{-i\omega_0 t} = 0. \tag{4}$$

Substituting relation (3) to (2), using condition (4) and keeping only resonance terms proportional to $e^{i\omega_0 t}$, we get well-known abridged equation for complex amplitude

$$\dot{a} = -\frac{\gamma}{2}a + \frac{3}{8}\frac{i\beta}{\omega_0}|a|^2 a. \tag{5}$$

Let us further introduce the real amplitude $R(t)$ and real phase $\varphi(t)$ with the help of next ratio

$$a(t) = R(t)e^{i\varphi(t)}.$$

We substitute expression for $a(t)$ in the abridged equation (5) and separate the real and imaginary parts. Then for real amplitude $R(t)$ and phase $\varphi(t)$ we obtain system of the differential equations:

$$\dot{R} = -\frac{\gamma}{2}R, \quad \dot{\varphi} = \frac{3\beta}{8\omega_0}R^2. \tag{6}$$

Solutions of these equations give dependence of real amplitude $R(t)$ and phase $\varphi(t)$ from time in an interval between impulses:

$$R(t) = R_n e^{-\gamma t/2}, \quad \varphi(t) = \frac{3\beta}{8\omega_0}R_n^2 \frac{1-e^{-\gamma t}}{\gamma} + \varphi_n, \tag{7}$$

where $R_n$ and $\varphi_n$ - initial amplitude and phase right after $n$-th impulse.

From equation (3) it is possible to find the expressions specifying dependences of coordinate $X(t)$ and speed $V(t)$ of oscillator from time:

$$X(t) = R(t)\cos[\omega_0 t + \varphi(t)],$$

$$V(t) = -\omega_0 R(t)\sin[\omega_0 t + \varphi(t)].$$

Let us substitute expressions for real amplitude $R(t)$ and phase $\varphi(t)$ (7) in last relationships and we shall find dependences of coordinate and speed of oscillator from time in an interval between impulses:

$$X(t) = R_n e^{-\gamma t/2} \cos(\omega_0 t + \frac{3\beta}{8\omega_0}R_n^2 \frac{1-e^{-\gamma t}}{\gamma} + \varphi_n),$$

$$V(t) = -\omega_0 R_n e^{-\gamma t/2} \sin(\omega_0 t + \frac{3\beta}{8\omega_0}R_n^2 \frac{1-e^{-\gamma t}}{\gamma} + \varphi_n). \tag{8}$$

To the moment of the beginning of $(n+1)$-th impulse the time equal to period of external influence $T$ has passed. Because of it coordinate and speed of oscillator are equal accordingly to $X(T)$ and $V(T)$. Owing to δ-shaped character of influence, right after impulse the coordinate does not change, and speed get the addition equal to amplitude of external influence $C$. Therefore for coordinate and speed right after $(n+1)$-th impulse from (8) it is reached [10]:

$$X_{n+1} = R_n e^{-\gamma T/2} \cos(\omega_0 T + \frac{3\beta}{8\omega_0} R_n^2 \frac{1-e^{-\gamma T}}{\gamma} + \varphi_n),$$

$$V_{n+1} = -\omega_0 R_n e^{-\gamma T/2} \sin(\omega_0 T + \frac{3\beta}{8\omega_0} R_n^2 \frac{1-e^{-\gamma T}}{\gamma} + \varphi_n) + C. \quad (9)$$

These expressions represent desired two-dimensional map which gives dependences of coordinate and speed of oscillator right after $(n+1)$-th impulse from its coordinate and speed right after $n$-th impulse. They are more convenient for presentation in complex form with help of new complex variable $Z$:

$$Z = \left[iX + \frac{V}{\omega_0}\right]\sqrt{\frac{3\beta}{8\omega_0}\frac{1-e^{-\gamma T}}{\gamma}}. \quad (10)$$

Using initial conditions

$$X(t=0) = X_n = R_n \cos\varphi_n,$$
$$V(t=0) = V_n = -\omega_0 R_n \sin\varphi_n,$$

and also expression for complex variable $Z_n$, we have got from (9):

$$Z_{n+1} = A + BZ_n \exp(i(|Z_n|^2 + \psi)), \quad (11)$$

where parameters $A$, $B$ and $\psi$ are defined by way of parameters of initial system

$$A = \frac{C}{\omega_0}\sqrt{\frac{3\beta}{8\omega_0}\frac{1-e^{-\gamma T}}{\gamma}}$$
$$B = e^{-\gamma T/2},$$
$$\psi = \omega_0 T.$$

The map (11) is called Ikeda map [11]. To complete the picture it is necessary to give a well-known illustration of chart of dynamical regimes for Ikeda map (Fig.1a). The chart of dynamical regimes is a diagram on the parameter plane where domains of qualitatively distinct regimes are indicated by colors. To depict such a chart one scans step by step an area of interest on the parameter plane. At each point associated with one pixel the discrete map is iterated. Then, the nature of a regime is analyzed after transient process and arrival to attractor, and the point is marked by an appropriate color.

Other graphic representation of complex dynamics of nonlinear multi-parameter maps is the chart of Lyapunov exponent [12-15]. For construction of such chart in each point of parameters space the value of Lyapunov exponent Λ is calculated and is coded according to it in grey gradation. To values of Λ, close to zero, the white color is corresponded; to negative values - the tint darker, than more absolute value of Λ. To positive Lyapunov exponents the black color is corresponded. Also white color designates points, in which the iterative process diverges. The Lyapunov exponent chart for Ikeda map is presented on Fig.1b.

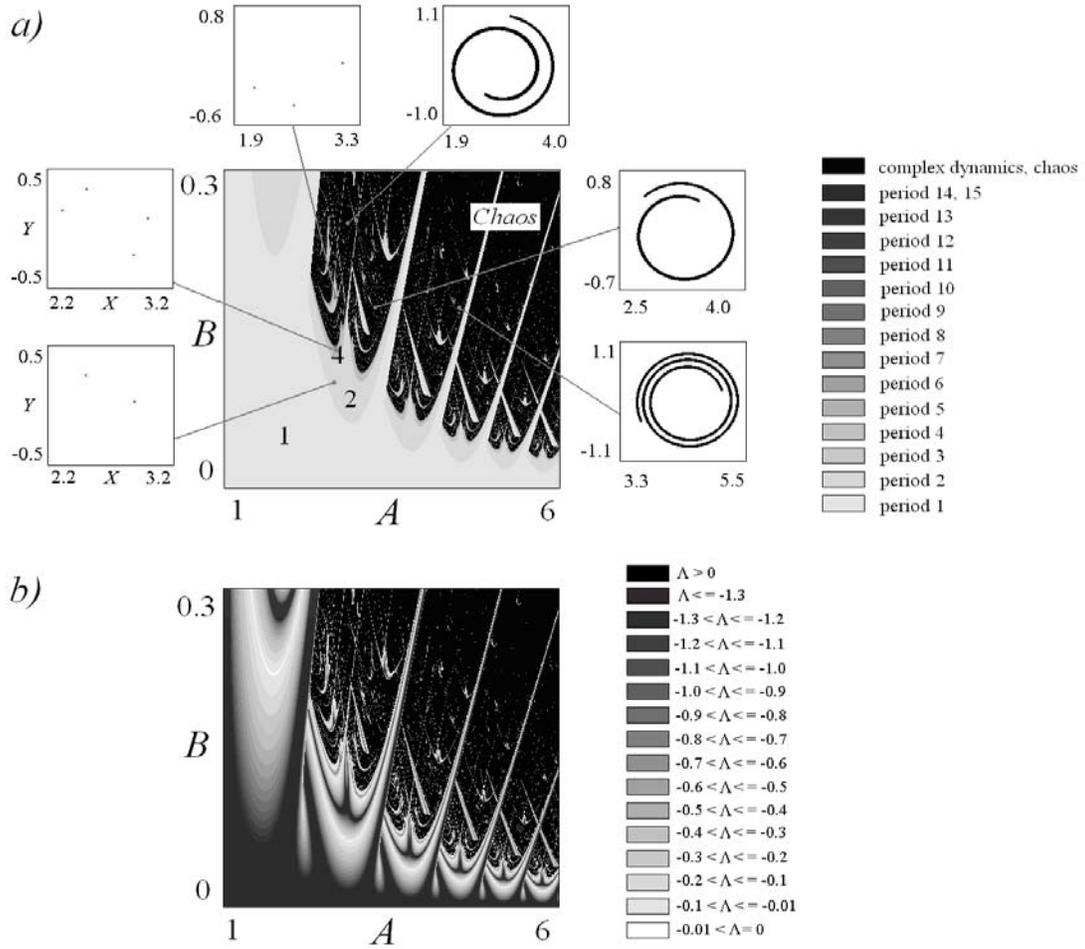

Figure 1: A chart of dynamical regimes with some phase portraits of attractor in indicated points on parameters plane (a) and chart of Lyapunov exponent Λ(b) for Ikeda map (11). With help of figures on chart of dynamical regimes the regions of existence of cycles with basic periods are marked.

Let now amplitudes of impulses modulate in a random way so, that $\Delta C_n$ - random addition to amplitude of $n$-th impulse:

$$\ddot{X} + \gamma \dot{X} + \omega_0^2 X + \beta X^3 = \sum (C + \Delta C_n) \delta(t - nT). \qquad (12)$$

For system (12) under action of impulses with random amplitude by analogy with (11) the following map is obtained:

$$Z_{n+1} = A\left(1 + \frac{\Delta C_n}{C}\right) + BZ_n \exp\left(i\left(|Z_n|^2 + \psi\right)\right). \tag{13}$$

It is possible to consider other way of insertion of fluctuations in investigated system. Let us suppose, that the duration of time intervals between impulses slightly changes about the average period $T$, so, that the duration of these intervals is $T + \Delta T_n$, where $\Delta T_n$ - small random value. Let us consider (that naturally for oscillator), that parameter $\gamma$ is much less than eigen frequency $\omega_0$. Then it is easily to see from (8), that as time $T + \Delta T_n$ passes the random addition to oscillator amplitude can be neglected in comparison with the addition to its complete phase. It means, that we come to the following form of Ikeda map with random influence

$$Z_{n+1} = A + BZ_n \exp\left(i\left(|Z_n|^2 + \psi + \omega_0 \Delta T_n\right)\right). \tag{14}$$

Illustration of noise effect on the Lyapunov exponent chart of Ikeda map with two above-mentioned ways of adding of noise is shown in Fig.2.

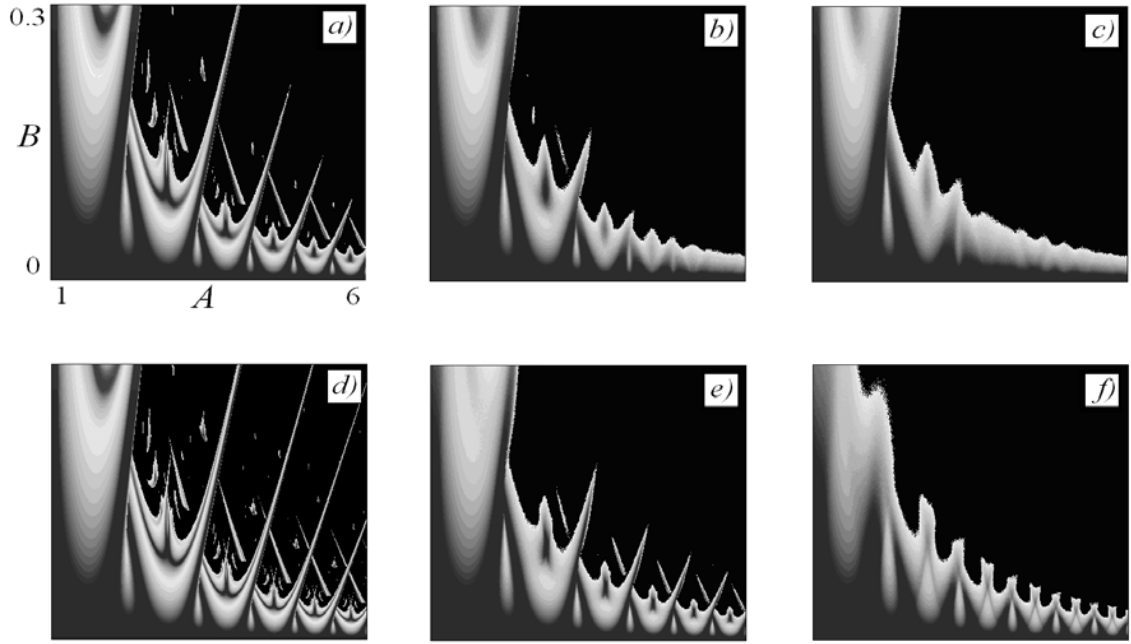

Figure 2: Illustration of noise influence at Ikeda map: Lyapunov charts for the map (13), where $\Delta C_n/C = \varepsilon \xi_n$, with different values of amplitude $\varepsilon = 0.01$ (a), 0.05 (b), 0.1 (c), and Lyapunov charts for the map (14), where $\omega_0 \Delta T_n = \varepsilon \xi_n$, noise amplitude $\varepsilon$ is changed: 0.1 (d), 1.0 (e), 3.0 (f). We suppose, that $\varepsilon$ - non-dimensional magnitude of modulation of the external influence amplitude (or period) and $\xi_n$ - random value.

Maps (13) and (14) are for the present two-dimensional (because $Z$ - complex variable, containing the real and imaginary parts). It is known, however, that in wide region of parameters values (the more precisely the more $A$) the Ikeda map admits a description with the help of one-dimensional map [9, 10]. Let us do the similar procedure for map (14). Following [10, 16] let us assign

$$Z = A + \widetilde{Z}, \qquad (15)$$

where $\widetilde{Z}$ is small addition. Let us substitute this expression in the left and right parts of the equation (14) taking into account smallness of $\widetilde{Z}$:

$$\widetilde{Z}_{n+1} = BA \exp\left(i\left(A^2 + 2A\operatorname{Re}\widetilde{Z}_n + \psi + \omega_0 \Delta T_n\right)\right). \qquad (16)$$

Having put further $X_n = 2A\operatorname{Re}\widetilde{Z}_n + A^2 + \psi$, we come to the following one-dimensional map with noise influence:

$$X_{n+1} = \lambda \cos(X_n + \omega_0 \Delta T_n) + \varphi, \qquad (17)$$

where the new parameters are used

$$\lambda = 2A^2 B, \quad \varphi = A^2 + \psi. \qquad (18)$$

Further we shall suppose, that $\omega_0 \Delta T_n = \varepsilon \xi_n$, where $\varepsilon$ is non-dimensional amplitude of modulation of the external influence period, and $\xi_n$ is random value.

The charts of dynamical regimes and Lyapunov exponent for cosine map (17) at the absence of external noise influence are given in a Fig.3. The range of changing of parameter $\varphi$ is chosen as ($-\pi/2$, $3\pi/2$) because of $2\pi$-periodicity of cosine function.

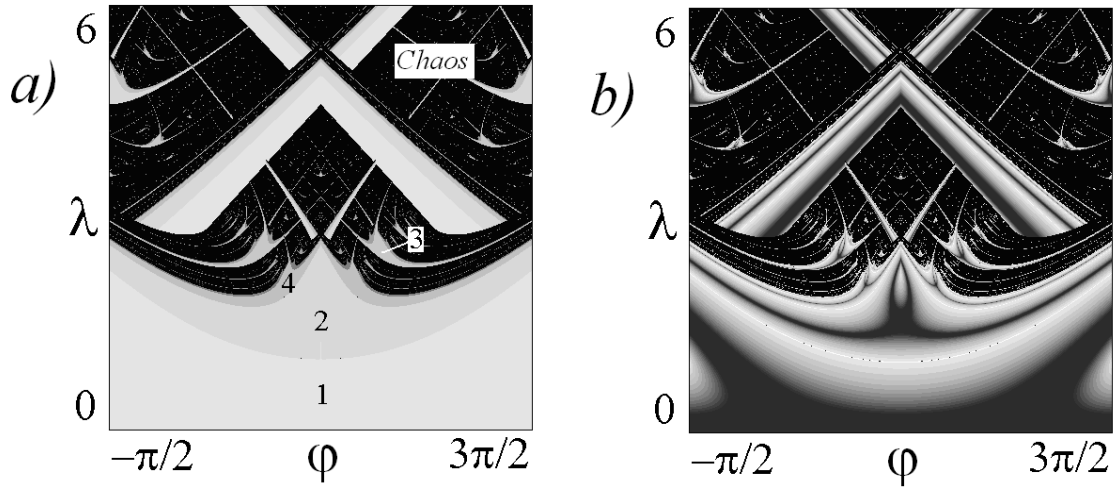

Figure 3: A chart of dynamical regimes (a) and chart of Lyapunov exponent $\Lambda$ (b), computing for cosine map (17) in the absence of noise. With help of Arabic numerals on chart of dynamical regimes the regions of existence of cycles with basic periods are indicated.

## 3. One-parametrical transition to chaos and scaling

The cosine map (17) in absence of fluctuations $X_{n+1} = \lambda \cos X_n + \varphi$ has a set of quadratic extrema. Therefore for such map the scenario of transition to chaos through period-doublings is characterized. At crossing the border of chaos on a typical route on a parameters plane ($\lambda$, $\varphi$) the classic cascade of doublings, described by regularities of Feigenbaum, will be observed. For example, at $\varphi = 0$ the Feigenbaum cascade is accumulated to a critical point $\lambda = 1.974133\ldots$ The illustration of scaling in system with noise on the bifurcation tree is shown in Fig.4. The self-similarity is illustrated by a series of figures, at that each subsequent figure of series represents the increased previous fragment. Thus scale on an axis of dynamical variable $X$ is recalculated in $\alpha_F$ relatively to a point $X=0$ - one of extrema of cosine map, and on an axis of control parameter $\lambda$ – in $\delta_F$ relatively to a critical point $1.974133\ldots$ According to the approach, stated in Introduction, at transition to deeper level of a hierarchical picture the initial amplitude of noise decreases at a constant $\mu_F = 6.61903\ldots$

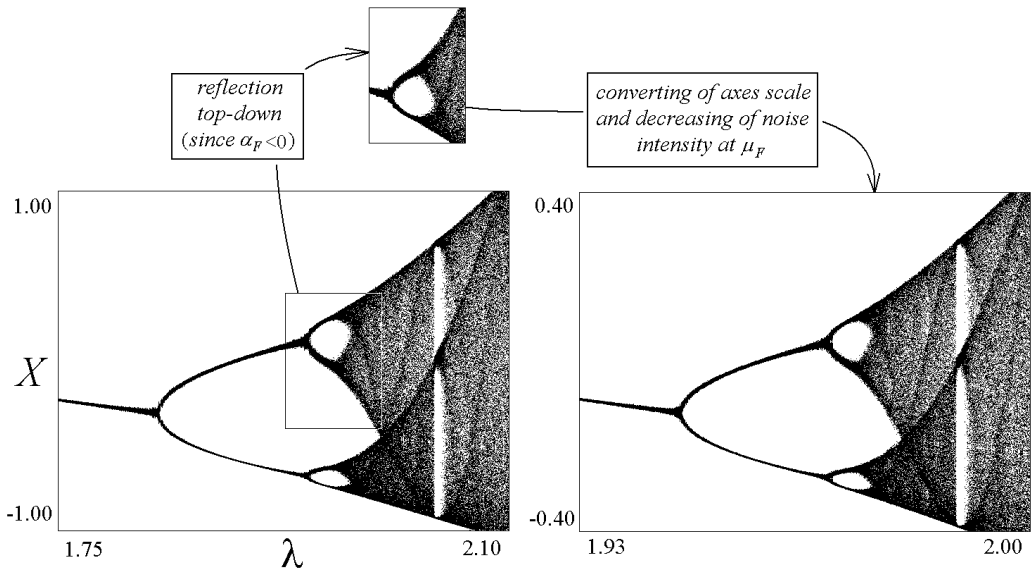

Figure 4: An illustration of Feigenbaum scaling property at bifurcation tree of cosine map. The initial noise intensity is equal to 0.003.

However, the map (17) is characterized by two parameters ($\lambda$, $\varphi$), therefore on this plane it is possible to observe also tricritical points. As we already marked in Introduction, for Feigenbaum lines they are "attribute" of maps having two and more quadratic extrema. In these points it is possible to reach also observing period-doubling cascades along lines, on which the maximum is mapped precisely in a minimum [17]. The reason of non-Feigenbaum character of convergence in this case is that twice iterated map under such condition has not quadratic extremum, but extremum of the fourth order. So for cosine map (17) at condition $\lambda = \pi - \varphi$ the quadratic maximum $X = 0$ is mapped precisely in a quadratic minimum $X = \pi$. Thus, along line $\lambda = \pi - \varphi$ on a parameters plane $(\lambda, \varphi)$ non-Feigenbaum cascade of period-doublings, accumulating to tricritical point $\lambda_T = 2.18603861533..$, $\varphi_T = 0.9555540392...$ [9], should be observed. Fig.5 demonstrates an illustration of scaling on the bifurcation tree for map (17) with typical for tricriticality constants $\delta_T = 7.284686217...$ along an axis of control parameter and $\alpha_T = -1.6903029714...$ along an axis of dynamical variable. At transition from one level of period-doublings to another the magnitude $\varepsilon$ decreases in $\mu_T = 8.2439...$ [18]. So, it is possible to see, that scaling in system with noise is well executed, that is the evidence of universality of scaling property concerning a concrete form of map with renormalization of noise intensity both for Feigenbaum, and for tricritical dynamics.

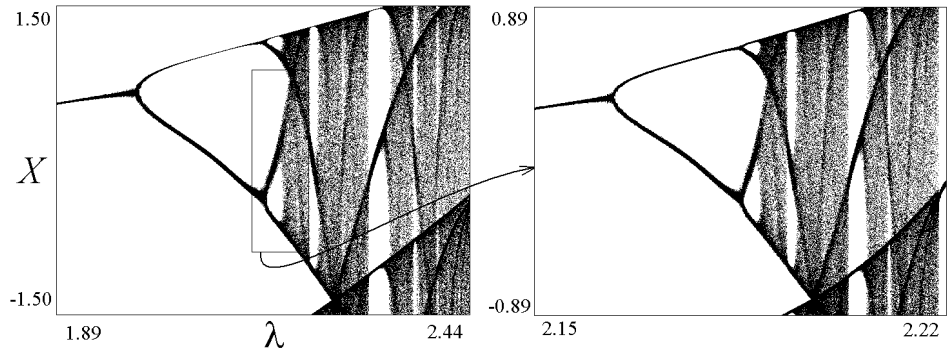

Figure 5: "Tricritical" scaling on bifurcation tree of cosine map (17) at the presence of noise with initial amplitude 0.01.

Separately it is necessary to mention that investigations on sensitivity of the obtained results to various types of noise influence were carried out. A binary noise, a noise with uniform distribution and a noise with Gauss distribution were generated. It turned out, that in all cases the property of self-similarity is realized, that indicates universality of scaling property also in relation to a kind of noise. For an illustration of all results of the given work for definiteness a noise influence with uniform distribution was chosen.

**4. Two-parametrical transition to chaos and scaling**

Let us turn now to the two-parametrical analysis. For systems without noise it assumes bifurcation analysis (construction of bifurcation lines on a parameters plane) or analysis of charts of dynamical regimes. For systems with noise both above-mentioned approaches are impossible. Therefore we use construction of above-mentioned Lyapunov exponent charts. The family of Lyapunov exponent charts for map (17) at various values of noise intensity $\varepsilon$ is shown in Fig.6. The range of changing of parame-ter $\varphi$ is from $\pi/2$ up to $\pi$, parameter $\lambda$ - from 1.5 up to 3.5.

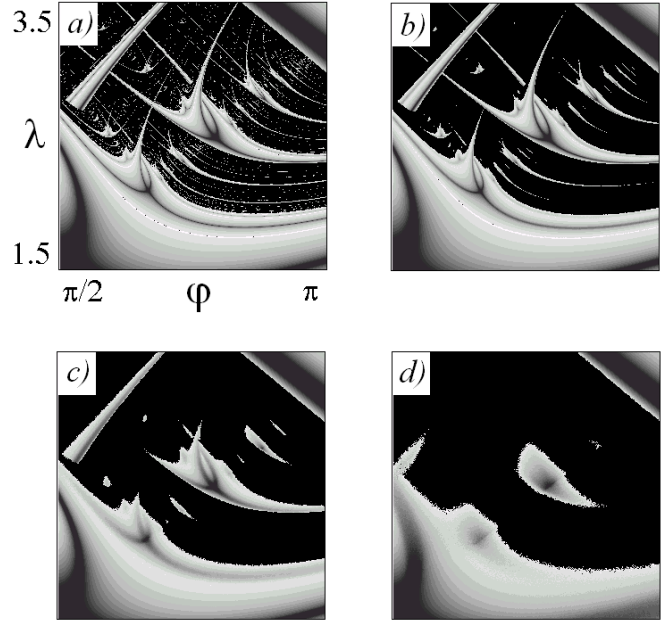

Figure 6: Family of Lyapunov charts for map (17) for various values of noise amplitude ε = 0.005 (a), 0.01 (b), 0.05 (c) and 0.2 (d).

In the case of small noise ε we see rather clear structure of Lyapunov space (Fig. 6a): the borders between certain values of Lyapunov exponent are well defined; the areas with zero Λ are sharply allocated; in chaos region the areas of periodic regimes are well visible. The noise of "average" intensity leads to disappearance of some regular regimes in the chaos area, though keeps general structure of a chart (Fig. 6b, c). At the further increasing of noise amplitude an almost complete disappearance of regular regimes is observed in the chaos region, the structure of a chart becomes more smeared, on all plane values of Lyapunov exponent are increased (Fig. 6d).

The presented charts of Lyapunov exponent are characterized by thin and complicated organization containing a set of small (in absence of noise indefinitely small) details. Most representative in this respect are tricritical points, in which vicinity the parameters plane is characterized by hierarchical organization submitted to property of self-similarity. Illustration of scaling requires, however, an introduction on a parameters plane the special scaling coordinate system. To base on the earlier obtained results, we shall slightly simplify map (17). For this purpose let us assume, that

$$X_n = \pi/2 + y_n - \varepsilon\xi_n. \tag{19}$$

In result we shall get

$$y_{n+1} = -\lambda \sin y_n + \varphi - \pi/2 + \varepsilon\xi_n. \tag{20}$$

Let us expand sine function up to the cubic term inclusive and then we shall introduce new variable and parameters according to correlations

$$X = y\sqrt{\lambda/6}, \quad a = (\varphi - \pi/2)\sqrt{\lambda/6}, \quad b = \lambda,$$

and amplitude of noise is normalized on the factor $\sqrt{\lambda/6}$. Then we come to cubic map

$$X_{n+1} = a - bX_n + X_n^3 + \varepsilon\xi_n. \tag{21}$$

The Lyapunov chart of this map in case $\varepsilon = 0$ is shown at the left of Fig. 7. It is possible to see, that above-mentioned approximations are not essential from the point of view of a chart's structure. (This concerns also the characteristics of tricritical dynamics, because such dynamics is a "rough" phenomenon in system with two quadratic extrema. The transition from cosine map (17) to cubic map (21) only slightly moves coordinates of two such essential extrema.)

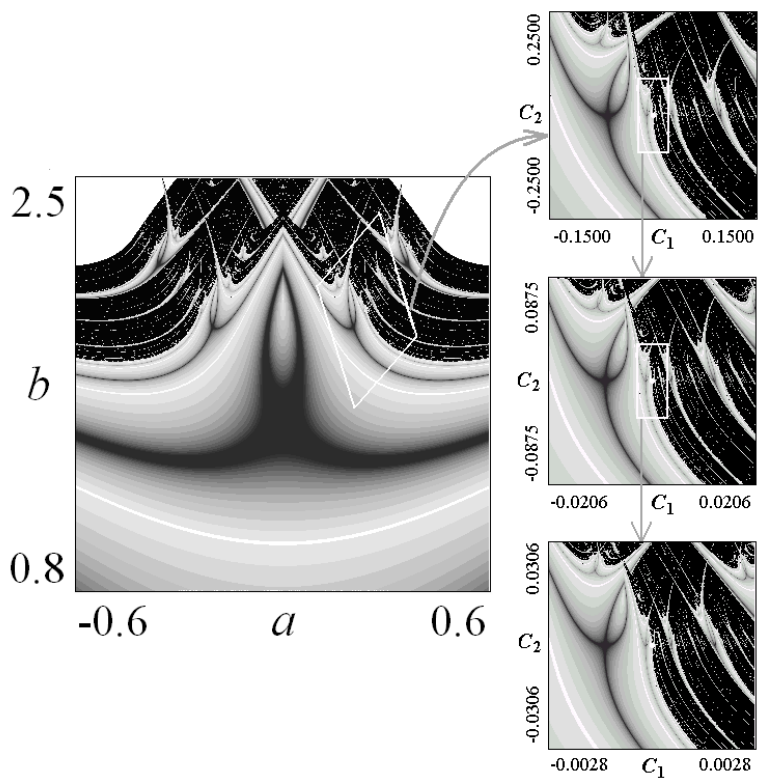

Figure 7: Scaling on Lyapunov exponent chart for map (21) in absence of noise.

The illustration of scaling property for map (21) on a parameters plane for a Lyapunov exponent chart is given in Fig.7. On large left fragment the coordinates axes are parameters of initial map ($a$, $b$); depicted parallelepiped (at which center the tricritical point is placed) is formed by coordinate lines $C_1 = \pm 0.15$ and $C_2 = \pm 0.25$. Here ($C_1$, $C_2$) are special coordinates, which are necessary to use for observa-tion of self-similarity. The connection between parameters of initial map ($a$, $b$) and "scaling coordinates" ($C_1$, $C_2$) is determined as follows [17, 19]:

$$a - a_T = 0.5998610 C_1 - 0.2192807 C_2, \quad b - b_T = C_1 + C_2. \quad (22)$$

Tricritical point $a_T$, $b_T$ has coordinates (0.2426987573.., 1.9513857778…).

The part of a parameters plane that got inside of mentioned parallelepiped is shown separately on the top right fragment already in "scaling coordinates", and then is twice reproduced with recalculation of scale on axis $C_1$ in factor $\delta_T$ and on axis $C_2$ - in factor $\alpha_T^2$ concerning the center of scaling – tricritical point. For observation of scaling at transition from one level to another the value of Lyapunov exponent need to be recalculated in 2 times in comparison with the previous fragment. According to this rule the color palette is changed.

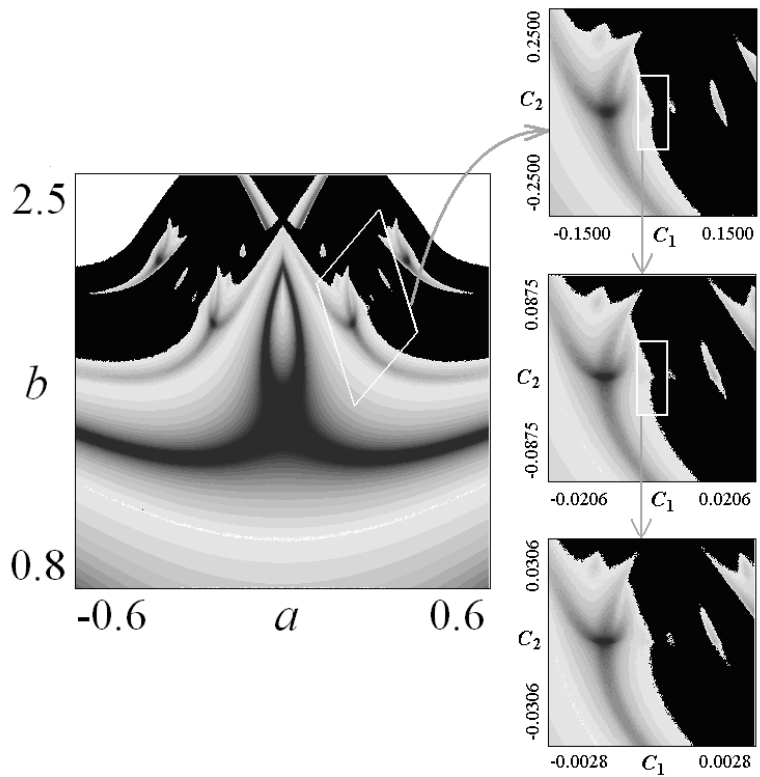

Figure 8: Scaling on Lyapunov exponent chart for cubic map (21) in the presence of noise with initial intensity $\varepsilon = 0.02$.

For observation of scaling properties for cubic map in the presence of noise (21) on a chart of Lyapunov exponent it is necessary to make the same recalculation, as well as in the case without noise, plus to reduce initial noise amplitude on each new fragment in comparison with previous one at $\mu_T = 8.2439…$ The appropriate illustrations for initial noise intensity $\varepsilon = 0.02$ are given in a Fig.8. It is easy to see, that each fragment with high accuracy repeats structure of previous fragment. That is an illustration of two-parametrical scaling in system with noise.

## 5. Conclusion

Thus nonlinear Duffing oscillator with impulse excitation and with random modulated amplitude or period of impulse sequence is convenient model for study of the critical phenomena at transition "order – chaos" in systems with noise. The maps obtained for this system, demonstrate both one-parametrical Feigenbaum scaling (but with an additional noise constant $\mu_F = 6.61903...$), and scaling on a plane of parameters (a Lyapunov exponent chart) with a noise constant $\mu_T = 8.2439...$

**Acknowledgements**

The work has been performed under support of Russian Science Support Foundation, Ministry of Education and Science of Russian Federation and CRDF (grant CRDF BRHE REC-006 SR-006-X1/BF5M06 Y3-P-06-07) and grant of the President of Russian Federation (MK-4162.2006.2).